\documentclass[oneside]{autart}
\usepackage[latin9]{inputenc}
\usepackage{color}
\usepackage{float}
\usepackage{mathrsfs}
\usepackage{amsmath}
\usepackage{amssymb}
\usepackage{graphicx}
\usepackage[unicode=true,pdfusetitle,
 bookmarks=true,bookmarksnumbered=false,bookmarksopen=false,
 breaklinks=false,pdfborder={0 0 0},pdfborderstyle={},backref=false,colorlinks=false]
 {hyperref}

\makeatletter

\floatstyle{ruled}
\newfloat{algorithm}{tbp}{loa}
\providecommand{\algorithmname}{Algorithm}
\floatname{algorithm}{\protect\algorithmname}

\newtheorem{notation}[thm]{Notation}

\makeatother

\begin{document}
\begin{frontmatter}

\title{Distributed Kalman Estimation with Decoupled Local Filters}

\author[gua,arg]{Dami\'{a}n Marelli}\ead{damian.marelli@newcastle.edu.au}, 
\author[dal]{Tianju Sui}\ead{suitj@mail.dlut.edu.cn}, 
\author[new]{Minyue Fu}\ead{minyue.fu@newcastle.edu.au}

\address[gua]{School of Automation, Guangdong University of Technology, Guangzhou, China.} 
\address[arg]{French-Argentinean International Center for Information and Systems Sciences, National Scientific and Technical Research Council, Rosario 2000, Argentina.}
\address[dal]{School of Control Science and Engineering, Dalian University of Technology, Dalian, China.}
\address[new]{School of Electrical Engineering and Computing, University of Newcastle, Callaghan, NSW 2308, Australia.}

\thanks{This work was supported by the Argentinean Agency for Scientific and Technological Promotion (PICT- 201-0985) and by the National Natural Science Foundation of China (Grant Nos. 61633014, 61803101 and U1701264).}

\begin{abstract}
We study a distributed Kalman filtering problem in which a number of nodes cooperate without central coordination to estimate a common state based on local measurements and data received from neighbors. This is typically done by running a local filter at each node using information obtained through some procedure for fusing data across the network. A common problem with existing methods is that the outcome of local filters at each time step depends on the data fused at the previous step. We propose an alternative approach to eliminate this error propagation. The proposed local filters are guaranteed to be stable under some mild conditions on certain global structural data, and their fusion yields the centralized Kalman estimate. The main feature of the new approach is that fusion errors introduced at a given time step do not carry over to subsequent steps. This offers advantages in many situations including when a global estimate in only needed at a rate slower than that of measurements or when there are network interruptions. If the global structural data can be fused correctly asymptotically, the stability of local filters is equivalent to that of the centralized Kalman filter. Otherwise, we provide conditions to guarantee stability and bound the resulting estimation error. Numerical experiments are given to show the advantage of our method over other existing alternatives.
\end{abstract}


\end{frontmatter}

\section{Introduction\label{sec:introduction}}

A networked system consists in a collection of nodes (or sub-systems),
connected via a communication network, executing certain processing
task~\cite{bullo2019lectures}. The processing is called distributed
if it is carried out by a cooperative strategy among nodes without
central coordination~\cite{luo2006distributed}. The design of distributed
methods aims at minimizing the amount of computation and communication
required by each node, as well as making these requirements scalable
in the number of nodes. Distributed methods are available for parameter
estimation~\cite{Xiao2006,Marelli2015}, Kalman filtering~\cite{Ribeiro2010},
control~\cite{Massioni2009,DAndrea2003}, optimization~\cite{Yang2019},
etc.

A Kalman filter gives the optimal maximum a posteriori estimation
of the state for linear systems with Gaussian noises. This is done
by alternating two steps called prediction and update. A major division
among distributed Kalman filtering methods is based on whether all
nodes estimate the full system state~\cite{Li2013}, or each node
only estimates a subset of the state variables~\cite{zhou2013coordinated,zhou2015controllability,farina2010moving,khan2008distributing,haber2013moving,sun2016dynamic}.
This work concerns with methods of the first type. Generally speaking,
all methods of this type assume that nodes know the state transition
equation. This permits that the prediction step is locally executed
at each node. The challenge then consists in how to distributedly
execute the update step. Most available methods do so by making use
of the information form of the Kalman filter. This requires computing
two quantities called the \emph{information vector} and \emph{information
matrix}, the former involving the fusion of measured signals at different
nodes and the latter involving the fusion of structural data of the
sub-systems. We broadly classify the available methods in two categories. 

In the first category the information vector and matrix are formed
by adding, using different communication schemes, partial components
from all nodes of the network. An early method was proposed in~\cite{Rao1991},
which requires full connectivity among all nodes. This restriction
was overcome in~\cite{Spanos2005} by using dynamic consensus~\cite{olfati2005consensus}
to fuse information across the network. The same method was refined
in~\cite{olfati2005distributed} by using different consensus stages
for fusing information vectors and matrices. In~\cite{casbeer2009distributed}
accuracy was improved, at the expense of extra communication, by adding
consensus sub-iterations between every two sample times. A variant
of this method was proposed~\cite{battistelli2014consensus} and
analyzed in~\cite{battistelli2016stability}, where two parallel
consensus stages are run for each, information vectors and matrices.
A different variant was proposed in~\cite{bai2011distributed}, where
a particular kind of dynamic consensus was used guaranteeing convergence
on the time-varying information vectors if certain assumptions are
met. In~\cite{das2015distributed,das2016consensus+}, fusion of information
vectors was done by representing them using a state-space model and
estimated them using a distributed Kalman filter of the second category.
In~\cite{hu2011diffusion} and~\cite{wang2017convergence} the fusion
scheme was complemented by using the covariance intersection method~\cite{julier2009general}
to fuse the outcomes of the prediction steps from each node. Finally,
in~\cite{wu2018distributed} fusion was done by using a message passing
algorithm, rather than a form of consensus, with the advantage of
finite-time convergence in the case of an acyclic communication network
graph.

In the second category, the fusion of information vectors and matrices
used in methods of the first category is complemented by fusion of
Kalman estimates. This approach was proposed in~\cite{olfati2007distributed}.
Its optimal design was studied in~\cite{olfati2009kalman} and its
performance analyzed in~\cite{khan2010connectivity}. A recent improvement
of this method was proposed in~\cite{li2019distributed}, by using
dynamic consensus to fuse information vectors and matrices. In~\cite{carli2008distributed}
the design was approached by proposing a particular structure, with
free parameters, which are optimized to minimize the estimation error.
A similar approach was later considered in~\cite{matei2012consensus}
using a more general structure and setup. Finally, in~\cite{cattivelli2010diffusion},
fusion of information vectors is eliminated and only Kalman estimates
are fused.

Broadly speaking, all methods from the two categories described above
require carrying out two kinds of data fusion. The first one aims
to fuse information associated with the parameters of the measurement
equation, typically to form the global information/covariance matrix.
We refer to this as \emph{structural data fusion}. The second one
aims to fuse information associated to the measurements locally acquired
at each node, typically to form the global information vector. We
refer to this as \emph{signal fusion}. In the case of time-invariant
measurement equations, structural data fusion needs to be done once,
possibly during initialization. Also, even in the time-varying case,
the rate of change of this data is typically slow, and can be easily
tracked using dynamic consensus with limited communications. In these
cases, structural data fusion can be done with negligible error. On
the other hand, the change of measurements across time steps is typically
much faster than that of the measurement equation. This requires a
signal fusion stage with more communications. 

A common property of the all the available methods described above
is that information fusion needs to be carried out at every Kalman
update step, because its result is needed for the subsequent Kalman
prediction and update steps. Due to the large communication demands
associated to this stage, it is often done approximately. The resulting
approximation error then propagates, in the sense that it affects
subsequent steps. This \emph{signal fusion error propagation} leads
to deviations between the estimates produced by the centralized Kalman
filter and those of their distributed counterparts, which accumulate
across time steps.

To overcome signal fusion error propagation, in this work we propose
an alternative method which avoids this drawback. In the proposed
method, each node runs a local estimator which does not require signal
fusion. Obviously, none of these local estimators can produce the
global Kalman estimate, since they only use local measurement information.
However, they have the property that the global Kalman estimate is
obtained by fusing their local estimates. In this way, the proposed
scheme avoids the aforementioned signal fusion error propagation problem.
For this reason, it is in our view a proper generalization of a Kalman
filter to a distributed setting. Apart from avoiding the accuracy
problems resulting from signal fusion error propagation, the proposed
scheme is advantageous in applications where a global estimate is
required at a rate slower than the one at which measurements are acquired.
This is because information fusion needs only be done at the slower
rate. Also, in the case of unreliable communications, where fusion
cannot be done during certain periods, the proposed scheme immediately
recovers without errors after communications resume.

An additional property of the proposed method is that, provided that
structural data fusion is accurately done, the stability of each local
estimator is equivalent to that of the centralized Kalman filter.
However, stability can be lost if the structural data fusion is done
with significant errors. We do a stability analysis in which we provide
a bound on the structural data fusion error that guarantees stability.
We also bound the difference between the distributed state estimate
and the centralized Kalman estimate due to both, structural data and
signal fusion errors.

The rest of the paper is organized as follows. In Section~\ref{sec:Problem-description}
we describe the research problem. In Section~\ref{sec:Overview-of-available}
we give an overview of the available approaches for distributed Kalman
filtering and point out their common drawback that motivates our work.
In Section~\ref{sec:Proposed-distributed-method} we introduce the
proposed distributed Kalman filtering scheme addressing the aforementioned
drawback. In Section~\ref{sec:Stability-and-accuracy} we present
our stability and accuracy analysis results and in Section~\ref{sec:Proofs-of-the}
we derive their proofs. In Section~\ref{sec:Numerical-experiments}
we give experimental evidence of our claims. Concluding remarks are
given in Section~\ref{sec:Conclusion}. For ease of readability,
the proofs of some auxiliary results appear in the Appendix.

\section{Problem description\label{sec:Problem-description}}
\begin{notation}
For a vector $x$, $\left\Vert x\right\Vert $ denotes its 2-norm
and for a matrix $X$, $\left\Vert X\right\Vert $ denotes its operator
norm. We use $\mathbb{S}_{N}\left(\mathbb{R}\right)\subset\mathbb{R}^{N\times N}$
to denote the set of real symmetric $N\times N$ matrices, and $\mathbb{P}_{N}\left(\mathbb{R}\right)\subset\mathbb{R}^{N\times N}$
to denote the set of real positive definite $N\times N$ matrices.
Also, $\mathrm{col}\left(x_{1},\cdots,x_{I}\right)$ denotes the column
vector formed by stacking the symbols (either vectors or matrices)
$x_{i}$, $i=1,\cdots,I$, and $\mathrm{diag}\left(x_{1},\cdots,x_{I}\right)$
denotes the diagonal matrix with the same symbols on its main diagonal.
We use $\mathbf{1}_{N}$ to denote the $N$-dimensional column vector
filled with ones, $\mathbf{I}_{N}$ to denote the $N$-dimensional
identity matrix and $\otimes$ to denote the Kronecker product. For
a symbol $\Xi^{i}$ we use the handy notation $\Xi^{-i}\triangleq\left(\Xi^{i}\right)^{-1}$
and $\Xi^{i\top}\triangleq\left(\Xi^{i}\right)^{\top}$.
\end{notation}
Consider a random vector sequence described by the following recursions
\begin{equation}
x_{t}=Ax_{t-1}+w_{t},\label{eq:SS1}
\end{equation}
where $\mathbb{R}^{N}\ni x_{0}\sim\mathcal{N}\left(\mu,P\right)$
and $w_{t}\sim\mathcal{N}\left(0,Q\right)$, with $P,Q\in\mathbb{P}_{N}\left(\mathbb{R}\right)$.
We assume that we have $I$ nodes acquiring measurements from $x_{t}$.
In order to model moving nodes, we assume that their associated measurement
equations are time-varying, i.e., at time step $t$, node~$i$ measures
\begin{equation}
y_{t}^{i}=C_{t}^{i}x_{t}+v_{t}^{i},\label{eq:measurement-eq}
\end{equation}
with $v_{t}^{i}\sim\mathcal{N}\left(0,R_{t}^{i}\right)$, $R_{t}^{i}\in\mathbb{P}_{M}\left(\mathbb{R}\right)$.
We assume that the set $\left\{ x_{0},w_{t},v_{t}^{i}:t\in\mathbb{N},i=1,\cdots,I\right\} $
is statistically mutually independent. 

Nodes are communicated via a consensus network. We assume that between
every two consecutive time steps $t$ and $t+1$, there are $K$ communication
cycles. In order to model a time-varying connection topology, at time
$t\in\mathbb{N}$ and cycle $k\in\{1,\cdots,K\}$, node~$i$ can
send messages to its neighbors $\mathcal{N}_{t,k}^{i}\subseteq\{1,\cdots,I\}$.
The communication link from node~$i$ to node~$j\in\mathcal{N}_{t,k}^{i}$
has gain $w_{t,k}^{j,i}$. The gains are such that the communication
graph is undirected, i.e., $w_{t,k}^{i,j}=w_{t,k}^{j,i}$. We also
assume that the adjacency matrix $W_{t,k}=\left[w_{t,k}^{i,j}\right]_{i,j=1}^{I}$
satisfies
\[
\lim_{K\rightarrow\infty}\lambda_{2}\left(W_{t,K}\times\cdots\times W_{t,1}\right)=0,
\]
where $\lambda_{2}\left(X\right)$ denotes the algebraic connectivity
of matrix $X$, i.e., the second largest eigenvalue. This guarantees
that, for any $x_{t,0}=\left[x_{t,1}^{i},\cdots,x_{t,1}^{I}\right]^{\top}\in\mathbb{R}^{I}$,
the sequence generated by $x_{t,k}=W_{t,k}x_{t,k-1}$ satisfies
\[
\lim_{k\rightarrow\infty}x_{t,k}=\mathbf{1}_{I}\otimes\frac{1}{I}\sum_{i=1}^{I}x_{0}^{i}.
\]

Writing~(\ref{eq:measurement-eq}) in block form we obtain
\begin{equation}
y_{t}=C_{t}x_{t}+v_{t},\label{eq:SS2}
\end{equation}
where $v_{t}\sim\mathcal{N}\left(0,R_{t}\right)$ and 
\begin{align*}
y_{t} & =\mathrm{col}\left(y_{t}^{1},\cdots,y_{t}^{I}\right),\\
v_{t} & =\mathrm{col}\left(v_{t}^{1},\cdots,v_{t}^{I}\right),\\
C_{t} & =\mathrm{col}\left(C_{t}^{1},\cdots,C_{t}^{I}\right),\\
R_{t} & =\mathrm{diag}\left(R_{t}^{1},\cdots,R_{t}^{I}\right).
\end{align*}
A research challenge consists in deriving a distributed method for
running a Kalman filter on the system~(\ref{eq:SS1})-(\ref{eq:SS2}).
As mentioned in Section~\ref{sec:introduction}, a number of method
are available for doing so. In Section~\ref{sec:Overview-of-available}
we give an overview of these methods and point out their common drawback.
In Section~\ref{sec:Proposed-distributed-method} we propose a method
which avoids this drawback.

\section{Overview of available distributed methods\label{sec:Overview-of-available}}

In this section we briefly summarize available approaches for distributed
Kalman filtering. Let 
\begin{align}
x_{t+1|t}^{i} & =Ax_{t|t}^{i},\label{eq:KF-est-prediction}\\
\Sigma_{t+1|t}^{i} & =A\Sigma_{t|t}^{i}A^{\top}+Q,\label{eq:KF-cov-prediction}\\
x_{t|t} & =\Sigma_{t|t}\left(\Sigma_{t|t-1}^{-1}x_{t|t-1}+C_{t}^{\top}R_{t}^{-1}y_{t}\right),\label{eq:KF-est-update}\\
\Sigma_{t|t} & =\left(\Sigma_{t|t-1}^{-1}+C_{t}^{\top}R_{t}^{-1}C_{t}\right)^{-1},\label{eq:KF-cov-update}
\end{align}
denote the centralized Kalman filter equations, where the update step
is expressed in information form, and $x_{t|s}^{i}$ and $\Sigma_{t|s}^{i}$
denote the approximations obtained at node~$i$. All methods assume
that the number $I$ of nodes is known at each node. Notice that it
is possible to compute $I$ in a distributed manner using the method
proposed in~\cite{shames2012distributed}. They also assume that
all nodes know $A$ and $Q$ and the initial values $x_{0|0}=\mu$
and $\Sigma_{0|0}=P$. Then, at time step $t$, given an update estimate/covariance
pair $x_{t|t}^{i}$, $\Sigma_{t|t}^{i}$, the Kalman prediction step
can be carried out at each node using~(\ref{eq:KF-est-prediction})-(\ref{eq:KF-est-update}).
The different methods differ in how the Kalman update step is carried
out. In Section~\ref{subsec:Distributed-Kalman-update} we describe
how this is done in the two method categories mentioned in Section~\ref{sec:introduction}.
Carrying out this step requires some form of data fusion across nodes.
In Section~\ref{subsec:Information-fusion-using} we describe the
most common options used for doing so. Finally, in Section~\ref{subsec:Common-drawback-of}
we comment on a common limitation of all available approaches.

\subsection{Distributed Kalman update step\label{subsec:Distributed-Kalman-update}}

Let
\begin{align}
\Psi_{t} & =C_{t}^{\top}R_{t}^{-1}C_{t}=\sum_{i=1}^{I}C_{t}^{i\top}R_{t}^{-i}C_{t}^{i},\label{eq:Psi}\\
\psi_{t} & =C_{t}^{\top}R_{t}^{-1}y_{t}=\sum_{i=1}^{I}C_{t}^{i\top}R_{t}^{-i}y_{t}^{i}.\label{eq:psi}
\end{align}
We refer to $\psi_{t}$ and $\Psi_{t}$ as the (global) signal and
structural data, respectively. In view of~(\ref{eq:psi}) and~(\ref{eq:Psi}),
the signal $\psi_{t}$ and structural data $\Psi_{t}$ can be made
available at each node using some kind of data fusion. The fusion
stage yields, at each node~$i$, estimates $\psi_{t}^{i}$ and $\Psi_{t}^{i}$
of $\psi_{t}$ and $\Psi_{t}$, respectively. The different available
methods depend on how, using $\psi_{t}^{i}$ and $\Psi_{t}^{i}$,
the update step~(\ref{eq:KF-est-update})-(\ref{eq:KF-cov-update})
is computed at each node. We describe below how this is done in the
aforementioned two categories:

\subsubsection{Consensus on global signal data}

Using again any form of data fusion, an approximation $\psi_{t}^{i}$
of $\psi_{t}$ can be obtained at each node~$i$. Using this approximation,
in~\cite{casbeer2009distributed,battistelli2014consensus,battistelli2016stability,bai2011distributed,hu2011diffusion,wang2017convergence}
$x_{t|t}^{i}$ is obtained using~(\ref{eq:KF-est-update}), i.e.,
\begin{equation}
x_{t|t}^{i}=\Sigma_{t|t}^{i}\left(\Sigma_{t|t-1}^{-i}x_{t|t-1}^{i}+\psi_{t}^{i}\right).\label{eq:CIF2}
\end{equation}
Alternatively, the Kalman gain 
\[
K_{t}=\Sigma_{t|t}C_{t}^{\top}R_{t}^{-1},
\]
is used in~\cite{Spanos2005,olfati2005distributed} to compute
\begin{align}
x_{t|t}^{i} & =x_{t|t-1}^{i}+K_{t}\left(y_{t}-C_{t}x_{t|t-1}^{i}\right)\nonumber \\
 & =x_{t|t-1}^{i}+\Sigma_{t|t}\left(C_{t}^{\top}R_{t}^{-1}y_{t}-C_{t}^{\top}R_{t}^{-1}C_{t}x_{t|t-1}^{i}\right)\nonumber \\
 & =x_{t|t-1}^{i}+\Sigma_{t|t}\left(\psi_{t}-\Psi_{t}x_{t|t-1}^{i}\right)\nonumber \\
 & \simeq x_{t|t-1}^{i}+\Sigma_{t|t}^{i}\left(\psi_{t}^{i}-\Psi_{t}^{i}x_{t|t-1}^{i}\right).\label{eq:OST1}
\end{align}

\subsubsection{Consensus on global signal data and estimates}

In order to help the estimates in~(\ref{eq:OST1}) to converge to
a common value, in~\cite{olfati2007distributed,olfati2009kalman,khan2010connectivity,li2019distributed},
an extra term penalizing inter-node mismatches is added. This leads
to
\begin{align}
x_{t|t}^{i} & =x_{t|t-1}^{i}+\Sigma_{t|t}^{i}\left(\psi_{t}^{i}-\Psi_{t}^{i}x_{t|t-1}^{i}\right)\nonumber \\
 & +D_{t}\sum_{j\in\mathcal{N}_{i}}\left(x_{t|t-1}^{j}-x_{t|t-1}^{i}\right),\label{eq:OST2}
\end{align}
where matrix $D_{t}$ is a free parameter that needs to be designed.
In particular, the choice $D_{t}=I-\Sigma_{t|t}^{i}\Psi_{t}^{i}$
is implicitly made in~\cite{li2019distributed}.

\subsection{Information fusion using consensus\label{subsec:Information-fusion-using}}

In this section we describe the different data fusion methods used
in the distributed Kalman filtering literature. These methods apply
to the fusion of both, global signal data $\psi_{t}$ and global structural
data $\Psi_{t}$. We describe then for fusing signal data. Its application
to the fusion of structural data is straightforward.

\subsubsection{Local fusion of neighbor data}

In~\cite{das2015distributed,das2016consensus+,olfati2007distributed,olfati2009kalman,khan2010connectivity,matei2012consensus,cattivelli2010diffusion},
$\psi_{t}^{i}$ is built by using only data from neighbor nodes. More
precisely, they assume that $K=1$, i.e., there is a single communication
cycle between consecutive time steps. Let
\begin{equation}
\mathring{\psi}_{t}^{i}=\left(C_{t}^{i}\right)^{\top}\left(R_{t}^{i}\right)^{-1}y_{t}^{i}.\label{eq:psi-breve}
\end{equation}
Then 
\[
\psi_{t}^{i}=I\sum_{j=1}^{I}w_{t,1}^{i,j}\mathring{\psi}_{t}^{i}.
\]

\subsubsection{Global fusion using consensus}

In~\cite{casbeer2009distributed,battistelli2014consensus,battistelli2016stability,carli2008distributed},
the fusion is done using $K>1$ consensus iterations, i.e., they run
the following recursions 
\begin{equation}
\psi_{t,k}^{i}=\sum_{j=1}^{I}w_{t,k}^{i,j}\psi_{t,k-1}^{j},\label{eq:psi-i}
\end{equation}
initialized by$\psi_{t,0}^{j}=I\mathring{\psi}_{t}^{i}$. The fused
data is then $\psi_{t}^{i}=\psi_{t,K}^{i}$, i.e., the one yield after
$K$ cycles.

\subsubsection{Local fusion using dynamic consensus}

In~\cite{Spanos2005,olfati2005distributed,bai2011distributed,li2019distributed},
fusion is done using dynamic consensus. More precisely, they assume
$K=1$ and the fused local data $\psi_{t}^{i}$ is computed by modifying
its previous value $\psi_{t-1}^{i}$ with an update term, i.e., 
\[
\psi_{t}^{i}=\sum_{j=1}^{I}w_{t,1}^{i,j}\left[\psi_{t-1}^{j}+I\mathring{\psi}_{t}^{j}-I\mathring{\psi}_{t-1}^{j}\right].
\]

\subsubsection{Global fusion using dynamic consensus}

The advantage of dynamic consensus is that it leads to an approximation
error $\psi_{t}-\psi_{t}^{i}$ that decreases as so does the rate
of change of $\psi_{t}$. Also, the advantage of using $K>1$ consensus
iterations is that it also permits reducing this error, at the expense
of extra communications. These two advantages can be readily combined
to increase accuracy as follows
\begin{equation}
\psi_{t,k}^{i}=\sum_{j=1}^{I}w_{t,k}^{i,j}\psi_{t,k-1}^{j},\label{eq:dyn-cons1}
\end{equation}
initialized by
\begin{equation}
\psi_{t,0}^{j}=\psi_{t-1,K}^{j}+I\mathring{\psi}_{t}^{j}-I\mathring{\psi}_{t-1}^{j},\label{eq:dyn-cons2}
\end{equation}
The fused data is then $\psi_{t}^{i}=\psi_{t,K}^{i}$. This is the
fusion method that we use in this work.

\subsection{Common drawback of all available methods\label{subsec:Common-drawback-of}}

The methods described above require running two fusion stages for
computing~(\ref{eq:Psi}) and~(\ref{eq:psi}). The first one computes
the global structural data $\Psi_{t}$. Since $\Psi_{t}$ is structural
data, it is often time-invariant or its change from one time step
to the next one in typically slow. In the former case, it can be readily
computed during initialization using some fusion mechanism. Otherwise,
we can track its slow evolution using dynamic consensus with a relatively
small number $K$ of consensus iterations. On the other hand, the
second consensus stage computes $\psi_{t}$. Since this quantity depends
on the measurements $y_{t}$, its change across time steps is typically
much faster than that of $\Psi_{t}$. This requires using consensus
with a larger value of $K$ to make an accurate estimate $\psi_{t}^{i}$
of $\psi_{t}$ available at each node. A common feature of the available
methods described above is that the signal fusion error incurred in
the estimation $\psi_{t}^{i}$ is carried over to the next time step.
This requires that the estimation of $\psi_{t}$ is accurately done
at each time step, using a large number of consensus iterations $K$,
even if an estimate $x_{t|t}$ is not required at that step. In the
next section we propose an alternative distributed method which avoids
this drawback.

\section{Proposed distributed method\label{sec:Proposed-distributed-method}}

In this section we describe the proposed distributed Kalman filtering
method. The covariance prediction and update steps are carried out
using~(\ref{eq:KF-cov-prediction}) and~(\ref{eq:KF-cov-update}),
as in the methods described in Section~\ref{sec:Overview-of-available}.
For the state estimate, suppose that 
\[
x_{t-1|t-1}=\sum_{i=1}^{I}\xi_{t-1|t-1}^{i},
\]
for some $\xi_{t-1|t-1}^{i}$, $i=1,\cdots,I$, which are only known
at node~$i$. We then have
\begin{align*}
x_{t|t} & =Ax_{t-1|t-1}+K_{t}\left(y_{t}-C_{t}Ax_{t-1|t-1}\right)\\
 & =\left(I-K_{t}C_{t}\right)Ax_{t-1|t-1}+K_{t}y_{t}\\
 & =\sum_{i=1}^{I}\left[\left(I-\Phi_{t}\right)A\xi_{t-1|t-1}^{i}+K_{t}^{i}y_{t}^{i}\right],
\end{align*}
where
\[
\Phi_{t}=K_{t}C_{t}.
\]
Letting $K_{t}^{\top}=\left[\left(K_{t}^{1}\right)^{\top},\cdots,\left(K_{t}^{I}\right)^{\top}\right]$,
where for each $i=1,\cdots,I$, the number of columns of $K_{t}^{i}$
equals the dimension of $y_{t}^{i}$, and 
\begin{equation}
\xi_{t|t}^{i}=\left(I-\Phi_{t}\right)A\xi_{t|t-1}^{i}+K_{t}^{i}y_{t}^{i},\label{eq:loc-filt}
\end{equation}
we obtain
\begin{equation}
x_{t|t}=\sum_{i=1}^{I}\xi_{t|t}^{i}.\label{eq:estimate}
\end{equation}

The above means that if we could distributedly compute the structural
data $\Phi_{t}$ and $K_{t}^{i}$, then each node could run the local
filter~(\ref{eq:loc-filt}) without needing to exchange information
with its neighbors unless an estimate of $x_{t|t}$ is needed at time
step $t$. We address the distributed computation of $\Phi_{t}$ and
$K_{t}^{i}$ below.

From the information form of the Kalman filter, we have
\[
K_{t}=\Sigma_{t|t}C_{t}^{\top}R_{t}^{-1}.
\]
Hence, $K_{t}^{i}$ can be readily computed at each node using
\begin{equation}
K_{t}^{i}=\Sigma_{t|t}\left(C_{t}^{i}\right)^{\top}\left(R_{t}^{i}\right)^{-1}.\label{eq:K}
\end{equation}
Also, 
\[
\Sigma_{t|t}=\left(I-\Phi_{t}\right)\Sigma_{t|t-1},
\]
leading to
\begin{align}
\Phi_{t} & =I-\Sigma_{t|t}\Sigma_{t|t-1}^{-1}\nonumber \\
 & =I-\Sigma_{t|t}\left(\Sigma_{t|t}^{-1}-\Psi_{t}\right)=\Sigma_{t|t}\Psi_{t}.\label{eq:Phi}
\end{align}
Hence, $\Phi_{t}$ can be locally computed at each node provided an
estimate of $\Psi_{t}$ is available.

The resulting method then requires a fusion stage to compute an estimate
of $\Psi_{t}$ at each node, and another one for computing $x_{t|t}$
using~(\ref{eq:estimate}). As we mentioned, we do fusion using the
dynamic consensus procedure~(\ref{eq:dyn-cons1})-(\ref{eq:dyn-cons2}).
We use $K_{\Psi}$ and $K_{x}$ to denote the number of consensus
iterations used to compute $\Psi_{t}$ and $x_{t|t}$, respectively.
The resulting method is summarized in Algorithm~\ref{alg:1}.

\begin{algorithm}
\textbf{Initialization:} We assume that, for each $i\in\{1,\cdots,I\}$,
node~$i$ knows $I$, $A$, $Q$ and $P$. Set
\[
\xi_{0|0}^{i}=\mu,\quad\Sigma_{0|0}=P\quad\text{and}\quad\Psi_{0}^{i}=0.
\]

\textbf{Main iterations:} At each $t\in\mathbb{N}$, we assume that
node~$i$ knows $C_{t}^{i}$, $R_{t}^{i}$ and $y_{t}^{i}$.
\begin{enumerate}
\item \textbf{Prediction: }
\begin{align}
\Sigma_{t|t-1}^{i} & =A\Sigma_{t-1|t-1}^{i}A^{\top}+Q,\label{eq:cov-prediction}\\
\xi_{t|t-1}^{i} & =A\xi_{t-1|t-1}^{i}.\label{eq:est-prediction}
\end{align}
\item \textbf{Structural data fusion:} For $k=1,\cdots,K_{\Psi}$, run
\begin{equation}
\Psi_{t,k}^{i}=\sum_{j=1}^{I}w_{t,k}^{i,j}\Psi_{t,k-1}^{j},\label{eq:consensus-Psi}
\end{equation}
initialized by
\[
\Psi_{t,0}^{j}=\Psi_{t-1}^{j}+I\mathring{\Psi}_{t}^{j}-I\mathring{\Psi}_{t-1}^{j},
\]
where
\begin{equation}
\mathring{\Psi}_{t}^{j}=C_{t}^{j\top}R_{t}^{-j}C_{t}^{j}.\label{eq:Psi-breve}
\end{equation}
Upon completion set $\Psi_{t}^{i}=\Psi_{t,K_{\Psi}}^{i}$. In the
time-invariant case, run this step only at $t=1$.
\item \textbf{Update:} 
\begin{align}
\Sigma_{t|t}^{i} & =\left(\Sigma_{t|t-1}^{-i}+\Psi_{t}^{i}\right)^{-1},\label{eq:cov-update}\\
K_{t}^{i} & =\Sigma_{t|t}^{i}C_{t}^{i\top}R_{t}^{-i},\label{eq:K-1}\\
\Phi_{t}^{i} & =\Sigma_{t|t}^{i}\Psi_{t}^{i},\label{eq:Phi-1}\\
\xi_{t|t}^{i} & =\left(I-\Phi_{t}^{i}\right)\xi_{t|t-1}^{i}+K_{t}^{i}y_{t}^{i}.\label{eq:est-update}
\end{align}
\item \textbf{Signal fusion:} If an estimate is required at $t$, then for
$k=1,\cdots,K_{x}$, run
\begin{equation}
x_{t|t,k}^{i}=\sum_{j=1}^{I}w_{t,k}^{i,j}x_{t|t,k-1}^{j},\label{eq:consensus-update}
\end{equation}
initialized by 
\[
x_{t|t,0}^{j}=x_{s|s}^{j}+I\xi_{t|t}^{j}-I\xi_{s|s}^{j},
\]
where $s$ is the previous time an estimate was required. Upon completion
set $\xi_{t|t}^{i}=x_{t|t}^{i}=x_{t|t,K_{x}}^{i}$.
\end{enumerate}
\caption{Proposed distributed Kalman filtering algorithm.}
\label{alg:1}
\end{algorithm}

\begin{rem}
In many applications, the structural data $\Psi_{t}$ is typically
either time-invariant or changes slowly with time in comparison with
$x_{t|t}$. We then typically use $K_{\Psi}$ much smaller than $K_{x}$.
A smaller $K_{\Psi}$ reduces the complexity of the algorithm while
allowing us to keep track of slow changes of the structural data. 
\end{rem}
\begin{rem}
Let $M_{i}$ denote dimension of the measurement vector $y_{t}^{i}$
at node~$i$ and $L_{i}=\max\left\{ M_{i},N\right\} $. The complexity
of Algorithm~1 is as follows: Each prediction/update step requires
$O\left(L_{i}^{2}N\right)$ multiplications, each structural data
fusion stage requires $O\left(L_{i}^{2}M_{i}\right)$ and each signal
fusion stage $O\left(N^{2}\right)$.
\end{rem}
Clearly, if structural data and signal fusions are done without errors,
every time $t$ signal fusion occurs, the estimate $x_{t|t}^{i}$
produced at each node~$i$ equals the centralized Kalman estimate
$x_{t|t}$. The question then naturally arises as to whether the linear
maps $\left(y_{t}\right)_{t\in\mathbb{N}}\mapsto\left(d_{t}^{i}\right)_{t\in\mathbb{N}}$,
where $d_{t}^{i}=\xi_{t|t}^{i}-\sum_{j=1}^{I}\xi_{t|t}^{j}$, are
stable for each $i=1,\cdots,I$. This is guaranteed by setting ${\color{blue}\xi_{t|t}^{i}}{\color{blue}=}x_{t|t}^{i}$
at the end of every signal fusion step. This requires running signal
fusion steps on a regular basis. However, if the dynamics of local
filters~(\ref{eq:est-prediction}),~(\ref{eq:est-update}) (equivalently~(\ref{eq:loc-filt}))
are stable, this requirements can be dropped. The following result
gives conditions guaranteeing this.
\begin{thm}
If $\Psi_{t}^{i}=\Psi_{t}$, for all $i=1,\cdots,I$ and $t\in\mathbb{N}$
(i.e., structural data fusion is done without errors), and the centralized
Kalman filter $\left(y_{t}\right)_{t\in\mathbb{N}}\mapsto\left(x_{t|t}\right)_{t\in\mathbb{N}}$
is stable, then the local filters $\left(y_{t}^{i}\right)_{t\in\mathbb{N}}\mapsto\left(\xi_{t|t}^{i}\right)_{t\in\mathbb{N}}$
are stable.
\end{thm}
\begin{pf}
Since $\Psi_{t}^{i}=\Psi_{t}$, it follows from~(\ref{eq:cov-update}),~(\ref{eq:Phi-1})
and~(\ref{eq:cov-prediction}) that $\Phi_{t}^{i}=\Phi_{t}.$ Therefore,
from~(\ref{eq:K-1}),~(\ref{eq:est-update}) and~(\ref{eq:est-prediction}),
that
\[
\xi_{t|t}^{i}=\left(I-\Phi_{t}\right)A\xi_{t-1|t-1}^{i}+\Sigma_{t|t}C_{t}^{i\top}R_{t}^{-i}y_{t}^{i}.
\]
Hence, the dynamics of each local filter are determined by the matrix
$\left(I-\Phi_{t}\right)A$. The result then follows since this is
also the matrix that determines the dynamics of the centralized Kalman
filter.$\hfill\qed$
\end{pf}
On the other hand, if errors are introduced at the structural data
fusion stage, they will affect local filter dynamics by introducing
errors in the recursions~(\ref{eq:cov-prediction})-(\ref{eq:est-update}).
This in turn raises a question about which error tolerance can be
allowed at the structural data fusion stage so as to preserve the
stability of the local filters $\left(y_{t}^{i}\right)_{t\in\mathbb{N}}\mapsto\left(\xi_{t|t}^{i}\right)_{t\in\mathbb{N}}$,
as well as that of the mismatch map $\left(y_{t}\right)_{t\in\mathbb{N}}\mapsto\left(x_{t|t}^{i}-x_{t|t}\right)_{t\in\mathbb{N}}$
between the estimates produced at each node and the centralized Kalman
one. We address these two questions in the next section.

\section{Stability and accuracy analysis\label{sec:Stability-and-accuracy}}

In this section we study the accuracy requirements in the structural
data fusion stage to guarantee the stability of local filters. We
also derive a bound on the mismatch between the estimates $x_{t|t}^{i}$
produced at each node and the centralized Kalman estimate $x_{t|t}$,
caused by errors introduced at both fusion stages.
\begin{notation}
We use $\tilde{\Psi}_{t}^{i}=\Psi_{t}^{i}-\Psi_{t}$ to denote the
error introduced at each node by the structural data fusion stage.
We use the same notation, e.g., $\tilde{\Sigma}_{t|s}^{i}=\Sigma_{t|s}^{i}-\Sigma_{t|s}$
and $\tilde{\Phi}_{t}^{i}=\Phi_{t}^{i}-\Phi_{t}$ for the resulting
errors introduced in the values of $\Sigma_{t|s}^{i}$\textup{ and
$\Phi_{t}^{i}$, respectively. We also use $\breve{\xi}_{t|s}^{i}$
to denote the value of $\xi_{t|s}^{i}$ that would result if no errors
were introduced at the structural data fusion stage, and define $\tilde{\xi}_{t|s}^{i}=\breve{\xi}_{t|s}^{i}-\xi_{t|s}^{i}$.}
\end{notation}
Let 
\[
\bar{\tilde{\psi}}=\sup_{\begin{subarray}{c}
t\in\mathbb{N}\\
1\leq i\leq I
\end{subarray}}\left\Vert \tilde{\Psi}_{t}^{i}\right\Vert ,
\]
be a bound on the error introduced at all structural data fusion stages.
Our first result states a sufficient condition on $\bar{\tilde{\psi}}$
to guarantee the stability of all local filters.
\begin{thm}
\label{thm:stab} Let $\bar{\sigma}=\sup_{t\in\mathbb{N}}\left\Vert \Sigma_{t|t}\right\Vert $
and
\begin{align*}
\bar{\gamma} & =\sup_{t\in N}\sum_{s=1}^{t}\left\Vert \left(I-\Phi_{t-1}\right)A\times\cdots\times\left(I-\Phi_{s}\right)A\right\Vert ,\\
\beta & =\mathrm{sol}_{b}\left\{ b+\log\bar{\sigma}\left\Vert A\right\Vert ^{2}\left\Vert Q^{-1}\right\Vert b=0\right\} .
\end{align*}
If the centralized Kalman filter $\left(y_{t}\right)_{t\in\mathbb{N}}\mapsto\left(x_{t|t}\right)_{t\in\mathbb{N}}$
is stable, and 
\begin{equation}
\bar{\tilde{\psi}}\leq\min\left\{ \bar{\sigma}^{-1}\left[1-\exp\left(-\frac{\beta}{\sqrt{N}}\right)\right],\bar{\gamma}^{-1}\left\Vert A\right\Vert ^{-1}\right\} ,\label{eq:stab-cond}
\end{equation}
then, the local filters $\left(y_{t}^{i}\right)_{t\in\mathbb{N}}\mapsto\left(\xi_{t|t}^{i}\right)_{t\in\mathbb{N}}$
are stable.
\end{thm}
\begin{rem}
Theorem~\ref{thm:stab} states that, if the error tolerance $\bar{\tilde{\psi}}$
of structural data fusion is smaller than the threshold given in~(\ref{eq:stab-cond}),
the stability of local filters is equivalent to that of the centralized
Kalman filter. Notice that, if measurement equations are time-invariant,
so is the structural data, i.e., $\Psi_{t}=\Psi$, for all $t\in\mathbb{N}$.
Hence, arbitrarily accurate structural data fusion can be guaranteed,
either during an initialization phase, or asymptotically at running
time. In this case, stability of local filters is simply equivalent
to that of the centralized Kalman filter. Also notice that the required
boundness of $\bar{\gamma}_{t}$ is equivalent to the stability of
the centralized Kalman filter.
\end{rem}
Let $\mathbf{x}_{t|t}=\mathbf{1}_{I}\otimes x_{t|t}$ denote a vector
with $I$ copies of the centralized Kalman estimate $x_{t|t}$, and
$\hat{\mathbf{x}}_{t|t}=\mathrm{col}\left(x_{t|t}^{1},\cdots,x_{t|t}^{I}\right)$
denote the vector of estimates produced by each node. Let also \textbf{$\check{\mathbf{x}}_{t|t}=\mathbf{1}_{I}\otimes\check{x}_{t|t}$},
where 
\[
\check{x}_{t|t}=\sum_{i=1}^{I}\xi_{t|t}^{i},
\]
denotes the estimate that would be obtained at all nodes if no error
were introduced a the signal fusion stage. Our second result bounds
the covariance of the error $\tilde{\mathbf{x}}_{t|t}\triangleq\mathbf{x}_{t|t}-\hat{\mathbf{x}}_{t|t}$.
This bound depends on two terms. The first one depends on the error
$\bar{\tilde{\psi}}$ introduced at the structural data fusion stage
and the second one depends on the error $\left\Vert \check{\mathbf{x}}_{t|t}-\hat{\mathbf{x}}_{t|t}\right\Vert $
introduced at the signal fusion stage.
\begin{lem}
\label{lem:implicit} Let $\bar{\upsilon}\left(\bar{\tilde{\psi}}\right)=\sqrt{N}\left|\log\left(1-\bar{\sigma}\bar{\tilde{\psi}}\right)\right|$.
If~(\ref{eq:stab-cond}) holds, then the following equation has at
least one solution
\begin{equation}
x=\frac{\bar{\sigma}\left\Vert A\right\Vert ^{2}x}{\bar{\sigma}\left\Vert A\right\Vert ^{2}+\left\Vert Q^{-1}\right\Vert ^{-1}e^{-x}}+\bar{\upsilon}\left(\bar{\tilde{\psi}}\right).\label{eq:implicit}
\end{equation}
\end{lem}
\begin{thm}
\label{thm:accu} Let $\mathfrak{y}_{t}^{i}=\mathrm{col}\left(\breve{\xi}_{t|t-1}^{i},\mathring{\psi}_{t}^{i}\right)$
with $\mathring{\psi}_{t}^{i}$ given by~(\ref{eq:psi-breve}) and
\begin{equation}
\bar{\mathfrak{y}}=\sup_{\begin{subarray}{c}
t\in\mathbb{N}\\
1\leq i\leq I
\end{subarray}}\left\Vert \mathcal{E}\left\{ \mathfrak{y}_{t}^{i}\mathfrak{y}_{t}^{i\top}\right\} \right\Vert ^{1/2}.\label{eq:difficult-bound}
\end{equation}
Suppose that~(\ref{eq:stab-cond}) holds, and let $\bar{\delta}\left(\bar{\tilde{\psi}}\right)$
denote the smallest solution of~(\ref{eq:implicit}). Then
\begin{equation}
\left\Vert \mathcal{E}\left\{ \tilde{\mathbf{x}}_{t|t}\tilde{\mathbf{x}}_{t|t}^{\top}\right\} \right\Vert ^{1/2}\leq\mathscr{E}\left(\bar{\tilde{\psi}}\right)+\mathcal{E}\left\{ \left\Vert \check{\mathbf{x}}_{t|t}-\hat{\mathbf{x}}_{t|t}\right\Vert ^{2}\right\} ^{1/2},\label{eq:main}
\end{equation}
where
\[
\mathscr{E}\left(\bar{\tilde{\psi}}\right)=NI\left(\frac{\bar{\gamma}\bar{\mathfrak{y}}}{1-\bar{\gamma}\bar{\tilde{\psi}}}\right)^{2}\left(\bar{\phi}^{2}\left(\bar{\tilde{\psi}}\right)+\bar{\tilde{\sigma}}^{2}\left(\bar{\tilde{\psi}}\right)\right),
\]
with
\begin{align*}
\bar{\phi}\left(\bar{\tilde{\psi}}\right) & =\left[\left(e^{\bar{\delta}\left(\bar{\tilde{\psi}}\right)}-1\right)\bar{\psi}+e^{\bar{\delta}\left(\bar{\tilde{\psi}}\right)}\bar{\tilde{\psi}}\right]\bar{\sigma},\\
\bar{\tilde{\sigma}}\left(\bar{\tilde{\psi}}\right) & =\left(e^{\bar{\delta}\left(\bar{\tilde{\psi}}\right)}-1\right)\bar{\sigma},
\end{align*}
and $\bar{\psi}=\sup_{t\in\mathbb{N}}\left\Vert \Psi_{t}\right\Vert $.
\end{thm}
\begin{rem}
The above result is stated in terms of the bound $\bar{\mathfrak{y}}$.
We give in Section~\ref{subsec:About-computing} details on how to
compute this bound.
\end{rem}
\begin{rem}
Notice that the first term $\mathscr{E}\left(\bar{\tilde{\psi}}\right)$
in~(\ref{eq:main}) depends only on the bound $\bar{\tilde{\psi}}$
of the structural data fusion error, and the second term is the signal
fusion error at each sample time $t$. These two errors are determined
by the numbers $K_{\Psi}$ and $K_{x}$ of consensus iterations used
on each fusion stage. Notice also that, for $\mathscr{E}\left(\bar{\tilde{\psi}}\right)$
to be bounded, so need to be $\bar{\mathfrak{y}}$ and $\bar{\psi}$.
\end{rem}

\section{Proofs of the main results\label{sec:Proofs-of-the}}

The proofs of Theorems~\ref{thm:stab} and~\ref{thm:accu} are given
in Section~\ref{subsec:proofs}. Deriving these results requires
certain mathematical background, which is introduced in three preceding
sections. In Section~\ref{subsec:The-Riemannian-metric} we introduce
a Riemannian metric on the differentiable manifold $\mathbb{P}_{N}\left(\mathbb{R}\right)$
of positive-definite matrices and state its properties. In Section~\ref{subsec:A-Hilbert-C-module}
we introduce a convenient algebraic structure on random vectors, namely,
a Hibert C$^{\star}$-module. Finally, in Section~\ref{subsec:perturbed-LTV}
we use this structure to characterize the output covariance of a perturbed
linear time-varying (LTV) system.

\subsection{A Riemannian metric on $\mathbb{P}_{N}\left(\mathbb{R}\right)$\label{subsec:The-Riemannian-metric}}

For a given $N\in\mathbb{N}$, the set $\mathbb{P}_{N}\left(\mathbb{R}\right)$
of positive-definite matrices can be considered as a differentiable
manifold inside $\mathbb{R}^{N\times N}$. We define the following
map $\delta:\mathbb{P}_{N}\left(\mathbb{R}\right)\times\mathbb{P}_{N}\left(\mathbb{R}\right)\rightarrow[0,\infty)$:
\begin{defn}
\cite[Chapter 6]{bhatia2009positive} For $P,Q\in\mathbb{P}_{N}\left(\mathbb{R}\right)$
we define 
\[
\delta\left(P,Q\right)=\left\Vert \log Q^{-1/2}PQ^{-1/2}\right\Vert _{\mathrm{F}}.
\]
\end{defn}
It is shown in~\cite[Chapter 6]{bhatia2009positive} that the map
$\delta$ is a Riemannian metric on $\mathbb{P}_{N}\left(\mathbb{R}\right)$.
This metric enjoys the properties given in the following proposition,
whose proof appears in the appendix.
\begin{prop}
\label{prop:rim-dist-1} For $P,Q\in\mathbb{P}_{N}\left(\mathbb{R}\right)$
and $R\in\mathbb{S}_{N}\left(\mathbb{R}\right)$: 
\begin{enumerate}
\item $\delta\left(P^{-1},Q^{-1}\right)=\delta\left(P,Q\right)$;
\item for any $W\in\mathcal{P}^{M}\left(\mathbb{R}\right)$ and $M\times N$
matrix $B$, we have 
\[
\delta\left(W+BPB^{\top},W+BQB^{\top}\right)\leq\frac{\alpha}{\alpha+\beta}\delta\left(P,Q\right),
\]
where $\alpha=\max\left\{ \left\Vert BPB^{\top}\right\Vert ,\left\Vert BQB^{\top}\right\Vert \right\} $
and $\beta=\left\Vert W^{-1}\right\Vert ^{-1}$;
\item $\left\Vert P-Q\right\Vert \leq\left(e^{\delta\left(P,Q\right)}-1\right)\min\left\{ \left\Vert P\right\Vert ,\left\Vert Q\right\Vert \right\} $.
\item If $\left\Vert P^{-1}\right\Vert \left\Vert R\right\Vert <1$, then
\[
\delta\left(P,P+R\right)\leq\sqrt{N}\left|\log\left(1-\left\Vert P^{-1}\right\Vert \left\Vert R\right\Vert \right)\right|.
\]
\end{enumerate}
\end{prop}

\subsection{A Hilbert C$^{\star}$-module of random vectors\label{subsec:A-Hilbert-C-module}}

A Hilbert C$^{\star}$-module is an algebraic structure that offers
an elegant and compact way to work with random vectors and their covariance
matrices. In this section we very briefly introduce the concepts needed
for our analysis. A general treatment of Hilbert C$^{\star}$-modules
can be found in~\cite{Lance1995}, and its application to covariance
matrices in~\cite[Section 3.3.1]{kailath2000linear}.

Let $x$ and $y$ be $N$-dimensional real random vectors. We define
the following $\mathbb{R}^{N\times N}$-valued inner product
\[
\left\langle x,y\right\rangle _{\star}=\mathcal{E}\left\{ xy^{\top}\right\} .
\]
This inner product induces the following norm on $N$-dimensional
random vectors
\begin{align*}
\left\Vert x\right\Vert _{\star} & =\left\Vert \left\langle x,x\right\rangle _{\star}\right\Vert ^{1/2}.
\end{align*}
It is shown in~\cite[Chapter 1]{Lance1995} that $\left\Vert x\right\Vert _{\star}$
is indeed a norm. This norm enjoys the following additional property,
whose proof appears in the appendix:
\begin{lem}
\label{lem:autocorr}Let $x$ and $y$ be random vectors of the same
dimension. Then
\[
\left\Vert \left\langle x,y\right\rangle _{\star}\right\Vert \leq\left\Vert x\right\Vert _{\star}\left\Vert y\right\Vert _{\star}.
\]
\end{lem}

\subsection{Output covariance of perturbed LTV systems\label{subsec:perturbed-LTV}}

In this section we use the Hilbert C$^{\star}$-module structure described
in Section~\ref{subsec:A-Hilbert-C-module} to bound the output covariance
of a perturbed LTV system. Consider the following LTV system
\begin{align}
x_{t} & =A_{t-1}x_{t-1}+u_{t},\label{eq:LTV1}\\
x_{0} & =0,\label{eq:LTV2}
\end{align}
with $u_{t}$ being a possibly colored and non-stationary vector random
process. Suppose we have a perturbed version $\hat{A}_{t}=A_{t}+\tilde{A}_{t}$
of the sequence $A_{t}$ and let $\hat{x}_{t}$ denote the sequence
generated by~(\ref{eq:LTV1})-(\ref{eq:LTV2}) when $A_{t}$ is replaced
$\hat{A}_{t}$. The following lemma gives a bound on the norm $\left\Vert \hat{x}_{t}\right\Vert _{\star}$,
in terms of the the non perturbed sequence $A_{t}$ and a measure
of the perturbation $\tilde{A}_{t}=\hat{A}_{t}-A_{t}$. Its proof
appears in the appendix.
\begin{lem}
\label{lem:LTV-bound}Let $A=\left(A_{t}\right)_{t\in\mathbb{Z}}$
and $\hat{A}=\left(\hat{A}_{t}\right)_{t\in\mathbb{Z}}$ be two sequences
of square matrices of the same dimension and $\tilde{A}_{t}=\hat{A}_{t}-A_{t}$.
Let 
\begin{align*}
\hat{x}_{t} & =\hat{A}_{t-1}\hat{x}_{t-1}+u_{t},\\
\hat{x}_{0} & =0.
\end{align*}
Let also $\mu=\sup_{t\in\mathbb{N}}\left\Vert \tilde{A}_{t}\right\Vert $,
$\bar{u}=\max_{1\leq s\leq t}\left\Vert u_{s}\right\Vert _{\star}$
and
\[
\gamma=\sup_{t\in Z}\sum_{s=1}^{t}\left\Vert A_{t-1}\times\cdots\times A_{s}\right\Vert .
\]
If $\mu\gamma<1$, then
\[
\left\Vert \hat{x}_{t}\right\Vert _{\star}\leq\frac{\gamma}{1-\gamma\mu}\bar{u}.
\]
\end{lem}

\subsection{Proofs of the main results\label{subsec:proofs}}

In this section we give the proofs of Theorems~\ref{thm:stab} and~\ref{thm:accu}.
We arrive to them through a sequence of lemmas, whose proofs appear
in the appendix. The first lemma gives a bound of the difference between
the ideal predicted covariance $\Sigma_{t|t}$ and its approximation
$\Sigma_{t|t}^{i}$ at node~$i$. This difference is measured using
the Riemannian metric $\delta$ introduced in Section~\ref{subsec:The-Riemannian-metric}.
\begin{lem}
\label{lem:riemmanian} If $\left\Vert \tilde{\Psi}_{t}^{i}\right\Vert <\left\Vert \Sigma_{t|t}\right\Vert ^{-1}$,
for all $t\in\mathbb{N}$ and $i\in\{1,\cdots,I\}$, then
\begin{multline*}
\delta\left(\Sigma_{t|t}^{i},\Sigma_{t|t}\right)\\
\leq\frac{\left\Vert A\right\Vert ^{2}\left\Vert \Sigma_{t-1|t-1}\right\Vert \delta\left(\Sigma_{t-1|t-1}^{i},\Sigma_{t-1|t-1}\right)}{\left\Vert A\right\Vert ^{2}\left\Vert \Sigma_{t-1|t-1}\right\Vert +\left\Vert Q^{-1}\right\Vert ^{-1}e^{-\delta\left(\Sigma_{t-1|t-1},\Sigma_{t-1|t-1}^{i}\right)}}\\
+\sqrt{N}\left|\log\left(1-\left\Vert \Sigma_{t|t}\right\Vert \left\Vert \tilde{\Psi}_{t}^{i}\right\Vert \right)\right|.
\end{multline*}
\end{lem}
The next lemma characterizes the approximation error $\tilde{\xi}_{t|t}^{i}$
as the output of a perturbed LTV system.
\begin{lem}
\label{lem:xi}For all $t\in\mathbb{N}$ and $i\in\{1,\cdots,I\}$,
\begin{equation}
\tilde{\xi}_{t|t}^{i}=\left(I-\Phi_{t}-\tilde{\Phi}_{t}^{i}\right)A\tilde{\xi}_{t-1|t-1}^{i}+\left[\begin{array}{cc}
\tilde{\Phi}_{t}^{i}, & \tilde{\Sigma}_{t|t}^{i}\end{array}\right]\mathfrak{y}_{t}^{i},\label{eq:perturbedLTV}
\end{equation}
where $\mathfrak{y}_{t}^{i}$ is defined as in Theorem~\ref{thm:accu}
and
\begin{align*}
\left\Vert \tilde{\Phi}_{t}^{i}\right\Vert  & \leq\left[\left(e^{\delta\left(\Sigma_{t|t}^{i},\Sigma_{t|t}\right)}-1\right)\left\Vert \Psi_{t}\right\Vert \right.\\
 & \left.+e^{\delta\left(\Sigma_{t|t}^{i},\Sigma_{t|t}\right)}\left\Vert \tilde{\Psi}_{t}^{i}\right\Vert \right]\left\Vert \Sigma_{t|t}\right\Vert ,\\
\left\Vert \tilde{\Sigma}_{t|t}^{i}\right\Vert  & \leq\left(e^{\delta\left(\Sigma_{t|t}^{i},\Sigma_{t|t}\right)}-1\right)\left\Vert \Sigma_{t|t}\right\Vert .
\end{align*}
\end{lem}
The following lemma gives a bound on the norm $\left\Vert \tilde{\xi}_{t|t}^{i}\right\Vert _{\star}$
of the approximation error $\tilde{\xi}_{t|t}^{i}$ at each node.
\begin{lem}
\label{lem:xi-bound} If~(\ref{eq:stab-cond}) holds, then
\begin{equation}
\left\Vert \tilde{\xi}_{t|t}^{i}\right\Vert _{\star}\leq\frac{\bar{\gamma}\bar{\mathfrak{y}}}{1-\bar{\gamma}\bar{\tilde{\psi}}}\sqrt{\bar{\phi}^{2}\left(\bar{\tilde{\psi}}\right)+\bar{\tilde{\sigma}}^{2}\left(\bar{\tilde{\psi}}\right)}.\label{eq:xi-bound}
\end{equation}
\end{lem}
We can now give the proofs of our main results.
\begin{pf}
{[}of Theorem~\ref{thm:stab}{]} This is an immediate consequence
of Lemma~\ref{lem:xi-bound}.$\hfill\qed$
\end{pf}
\begin{pf}
{[}of Lemma~\ref{lem:implicit}{]} Let 
\[
a=\frac{\left\Vert Q^{-1}\right\Vert ^{-1}}{\bar{\sigma}\left\Vert A\right\Vert ^{2}},\qquad b=\bar{\upsilon}\left(\bar{\tilde{\psi}}\right),
\]
and
\[
f(x)=\frac{a}{b}\left(x-b\right),\qquad g(x)=e^{x}.
\]
Equation~(\ref{eq:implicit}) can then be rewritten as
\begin{equation}
f(x)=g(x).\label{eq:aux}
\end{equation}
Since $f$ is affine and $g$ convex,~(\ref{eq:aux}) has either
zero, one or two solutions. In order for it to have a single solution,
we must have 
\[
\frac{a}{b}=f^{\prime}(x)=g^{\prime}(x)=e^{x}.
\]
Replacing the above into~(\ref{eq:aux}) we obtain
\[
\log\frac{a}{b}=b
\]
or equivalently, $b=\beta$. It then follows that~(\ref{eq:implicit})
has at least one solution if $b\leq\beta$. It is straightforward
to verify that the latter is implied by~(\ref{eq:stab-cond}) and
the result follows.$\hfill\qed$
\end{pf}
\begin{pf}
{[}of Theorem~\ref{thm:accu}{]} We have
\begin{align}
 & \left\Vert \mathcal{E}\left\{ \tilde{\mathbf{x}}_{t|t}\tilde{\mathbf{x}}_{t|t}^{\top}\right\} \right\Vert ^{1/2}\nonumber \\
= & \left\Vert \mathcal{E}\left\{ \left(\mathbf{x}_{t|t}-\hat{\mathbf{x}}_{t|t}\right)\left(\mathbf{x}_{t|t}-\hat{\mathbf{x}}_{t|t}\right)^{\top}\right\} \right\Vert ^{1/2}\nonumber \\
\leq & \mathcal{E}\left\{ \mathrm{Tr}\left\{ \left(\mathbf{x}_{t|t}-\hat{\mathbf{x}}_{t|t}\right)\left(\mathbf{x}_{t|t}-\hat{\mathbf{x}}_{t|t}\right)^{\top}\right\} \right\} ^{1/2}\nonumber \\
= & \mathcal{E}\left\{ \sum_{i=1}^{I}\left\Vert x_{t|t}-\hat{x}_{t|t}^{i}\right\Vert ^{2}\right\} ^{1/2}\nonumber \\
= & \mathcal{E}\left\{ \sum_{i=1}^{I}\left\Vert x_{t|t}-\check{x}_{t|t}+\check{x}_{t|t}-\hat{x}_{t|t}^{i}\right\Vert ^{2}\right\} ^{1/2}\nonumber \\
\leq & \sqrt{I}\mathcal{E}\left\{ \left\Vert x_{t|t}-\check{x}_{t|t}\right\Vert ^{2}\right\} ^{1/2}+\mathcal{E}\left\{ \sum_{i=1}^{I}\left\Vert \check{x}_{t|t}-\hat{x}_{t|t}^{i}\right\Vert ^{2}\right\} ^{1/2}.\label{eq:aux-3}
\end{align}
Now, using Lemma~\ref{lem:xi-bound},
\begin{multline}
\mathcal{E}\left\{ \left\Vert x_{t|t}-\check{x}_{t|t}\right\Vert ^{2}\right\} =\mathcal{E}\left\{ \left\Vert \sum_{i=1}^{I}\xi_{t|t}-\hat{\xi}_{t|t}^{i}\right\Vert ^{2}\right\} \\
\leq\sum_{i=1}^{I}\mathcal{E}\left\{ \left\Vert \tilde{\xi}_{t|t}^{i}\right\Vert ^{2}\right\} =\sum_{i=1}^{I}\mathcal{E}\left\{ \mathrm{Tr}\left\{ \tilde{\xi}_{t|t}^{i}\tilde{\xi}_{t|t}^{i\top}\right\} \right\} \\
\leq N\sum_{i=1}^{I}\left\Vert \mathcal{E}\left\{ \tilde{\xi}_{t|t}^{i}\tilde{\xi}_{t|t}^{i\top}\right\} \right\Vert =N\sum_{i=1}^{I}\left\Vert \tilde{\xi}_{t|t}^{i}\right\Vert _{\star}^{2}\\
\leq NI\left(\frac{\bar{\gamma}\bar{\mathfrak{y}}}{1-\bar{\gamma}\bar{\tilde{\psi}}}\right)^{2}\left(\bar{\phi}^{2}\left(\bar{\tilde{\psi}}\right)+\bar{\tilde{\sigma}}^{2}\left(\bar{\tilde{\psi}}\right)\right).\label{eq:aux-4}
\end{multline}
The result then follows by putting~(\ref{eq:aux-4}) into~(\ref{eq:aux-3}),
and noticing that
\[
\mathcal{E}\left\{ \sum_{i=1}^{I}\left\Vert \check{x}_{t|t}-\hat{x}_{t|t}^{i}\right\Vert ^{2}\right\} =\mathcal{E}\left\{ \left\Vert \check{\mathbf{x}}_{t|t}-\hat{\mathbf{x}}_{t|t}\right\Vert ^{2}\right\} .
\]
$\hfill\qed$
\end{pf}

\section{About computing $\bar{\mathfrak{y}}$\label{subsec:About-computing}}

Our first step consists in characterizing $\mathfrak{y}_{t}^{i}$
as the output of a state-space model. This is done by defining
\[
\mathfrak{x}_{t}^{i}=\left[\begin{array}{c}
x_{t}\\
\breve{\xi}_{t|t}^{i}
\end{array}\right]\qquad\text{and}\qquad\mathfrak{e}_{t}^{i}=\left[\begin{array}{c}
w_{t}\\
v_{t}^{i}
\end{array}\right].
\]
We can then write
\begin{align}
\mathfrak{x}_{t}^{i} & =F_{t}^{i}\mathfrak{x}_{t-1}^{i}+G_{t}^{i}\mathfrak{e}_{t}^{i},\label{eq:SSA1}\\
\mathfrak{y}_{t}^{i} & =H_{t}^{i}\mathfrak{x}_{t}^{i}+E_{t}^{i}\mathfrak{e}_{t}^{i},\label{eq:SSA2}
\end{align}
with
\begin{align*}
F_{t}^{i} & =\left[\begin{array}{cc}
A & 0\\
K_{t}^{i}C_{t}^{i}A & \left(I-\Phi_{t}\right)A
\end{array}\right],\\
G_{t}^{i} & =\left[\begin{array}{cc}
I & 0\\
K_{t}^{i}C_{t}^{i} & K_{t}^{i}
\end{array}\right],\;H_{t}^{i}=\left[\begin{array}{cc}
0 & A\\
\mathring{\Psi}_{t}^{i} & 0
\end{array}\right],\;E_{t}^{i}=\left[\begin{array}{cc}
0 & 0\\
0 & C_{t}^{i\top}R_{t}^{-i}
\end{array}\right],
\end{align*}
and $\mathring{\Psi}_{t}^{i}$ given by~(\ref{eq:Psi-breve}). 

Using the model~(\ref{eq:SSA1})-(\ref{eq:SSA2}) we obtain the covariance
of $\mathfrak{y}_{t}^{i}$ as follows
\begin{align*}
\mathcal{E}\left\{ \mathfrak{x}_{t}^{i}\mathfrak{x}_{t}^{i\top}\right\}  & =\Pi_{0,t}^{i}\left[\begin{array}{cc}
P_{0} & 0\\
0 & 0
\end{array}\right]\Pi_{0,t}^{i\top}\\
 & +\sum_{s=1}^{t}\Pi_{s,t}^{i}G_{s}^{i}\left[\begin{array}{cc}
Q & 0\\
0 & R_{t}^{i}
\end{array}\right]G_{s}^{i\top}\Pi_{s,t}^{i\top},\\
\mathcal{E}\left\{ \mathfrak{y}_{t}^{i}\mathfrak{y}_{t}^{i\top}\right\}  & =H_{t}^{i}\mathcal{E}\left\{ \mathfrak{x}_{t}^{i}\mathfrak{x}_{t}^{i\top}\right\} H_{t}^{i\top}+E_{t}^{i}\left[\begin{array}{cc}
Q & 0\\
0 & R_{t}^{i}
\end{array}\right]E_{t}^{i\top},
\end{align*}
where $\Pi_{s,t}^{i}=F_{t}\times\cdots\times F_{s+1}$. We can then
readily compute the bound $\bar{\mathfrak{y}}$ by putting the above
into~(\ref{eq:difficult-bound}).

\section{Numerical experiments\label{sec:Numerical-experiments}}

In this section we evaluate the performance of our method. For comparison
we use one method form each of the two categories described in Section~\ref{sec:introduction}.
For the first category we consider the method proposed in~\cite{casbeer2009distributed}.
We refer to it as Algorithm~A. For the second category we consider
the method recently proposed in~\cite{li2019distributed}, which
we refer to as Algorithm~B.

For evaluation we use a randomly generated time-invariant system of
order $N=10$. Matrices $A$ and $Q$ have spectral radii $\rho(A)=0.999$
and $\rho(Q)=1$, respectively. Also, measurements are one-dimensional,
i.e., $M=1$, with $C^{i}\sim\mathcal{N}\left(0,\mathbf{I}_{N}\right)$
and $R^{i}=10r^{2}+0.1$, with $r\sim\mathcal{N}\left(0,1\right)$.
Nodes are connected via a time-invariant network with ring topology,
whose gains are given by
\[
w_{t,k}^{i,j}=\begin{cases}
0.5, & i=j,\\
0.25, & \left|\mod\left(i-j+1,I\right)-1\right|=1,\\
0, & \text{otherwise}.
\end{cases}
\]
This results in an algebraic connectivity of $\lambda_{2}=0.9891$.

As performance index we use the estimation mismatch error defined
as
\begin{align*}
e^{2} & =\frac{1}{T}\sum_{t=1}^{T}e_{t}^{2}\quad\text{with}\quad e_{t}^{2}=\frac{1}{I}\sum_{i=1}^{I}\left\Vert x_{t|t}^{i}-x_{t|t}\right\Vert ^{2},
\end{align*}
where $x_{t|t}$ denotes the centralized Kalman estimate.

\begin{figure}
\centering{}\includegraphics[width=0.7\columnwidth]{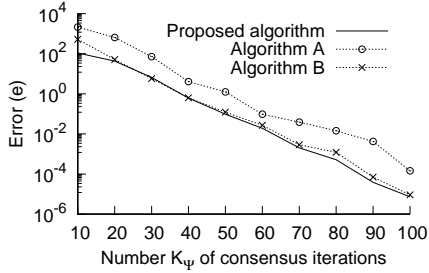}\caption{Error vs number of consensus iterations for structural data fusion.}
\label{fig:fusion-struct}
\end{figure}

\begin{figure}
\centering{}\includegraphics[width=0.7\columnwidth]{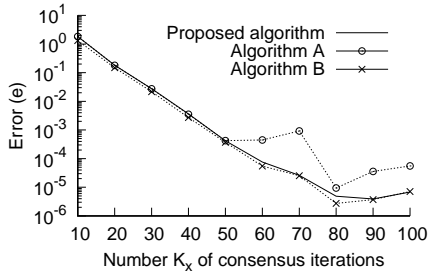}\caption{Error vs number of consensus iterations for signal fusion.}
\label{fig:fusion-signal}
\end{figure}

In the first experiment we evaluate the performance when errors appear
in the structural and signal fusion stages. In Figure~\ref{fig:fusion-struct}
we show the effect produced by an approximation error in fusing structural
data. To this end we use $K_{x}=100$ consensus iterations for signal
fusion and show the mismatch error as a function of the number $K_{\Psi}$
of consensus cycles used for structural fusion. We see that Algorithm~B
and the proposed one performs similarly, with a noticeable advantage
over Algorithm~A. In Figure~\ref{fig:fusion-signal} we use $K_{\Psi}=100$
iterations for structural fusion and show the mismatch error as a
function of the number $K_{x}$ of consensus iterations used for signal
fusion. We again see that Algorithm~B and the proposed one performs
similarly, with certain advantage over Algorithm~A for large values
of $K_{x}$. We conclude that, when there are no network interruptions,
the proposed algorithm performs similarly to the best available ones.

\begin{figure}
\centering{}\includegraphics[width=0.7\columnwidth]{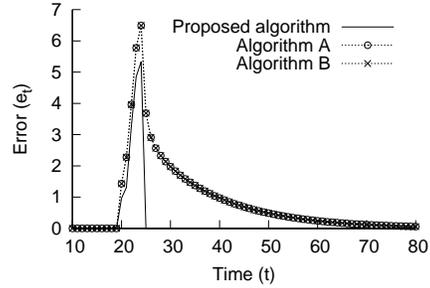}\caption{Performance under a network interruption.}
\label{fig:markov-single}
\end{figure}

\begin{figure}
\centering{}\includegraphics[width=0.7\columnwidth]{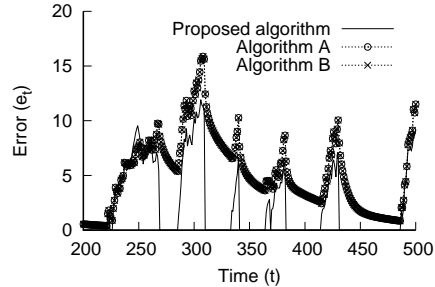}\caption{Performance under network interruptions following a Gilbert-Elliott
model.}
\label{fig:markov-many}
\end{figure}

As mentioned, the advantage of the proposed method is that errors
in signal fusion do not carry over across time steps. This can be
seen in Figure~\ref{fig:markov-single}, where we simulate a network
interruption from sample times $t=20$ to $t=25$. We use $K_{x}=K_{\Psi}=100$.
We see that, while the proposed algorithm gives an accurate estimate
as soon as connectivity is restored, Algorithms~A and~B require
several time steps to do so. In Figure~\ref{fig:markov-many} we
show the performance of the algorithms when network availability follows
a symmetric Gilbert-Elliott model~\cite{gilbert1960cbn,elliott1963eer}
with transition probability $p=0.05$. We see how, while the proposed
algorithm is always able to produce an accurate estimate as soon as
network connectivity is available, Algorithms~A and~B are not able
to produce accurate estimates during certain long time periods.

\begin{figure}
\centering{}\includegraphics[width=0.7\columnwidth]{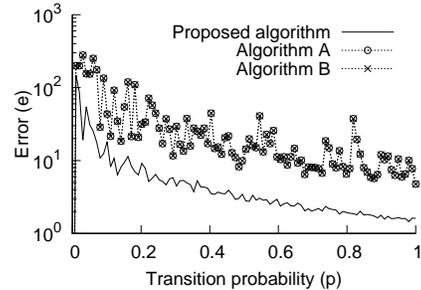}\caption{Error vs transition probability $p$.}
\label{fig:trans-prob}
\end{figure}

In Figure~\ref{fig:trans-prob} we show the mismatch error, as a
function of the transition probability $p$. We see that, in this
case, the proposed algorithm has a significant advantage over its
rivals.

\section{Conclusion\label{sec:Conclusion}}

We proposed a novel approach for distributed Kalman filtering. The
essential difference with existing approaches is that, provided certain
global structural data is available at each node, local filters do
not require data fusion across the network. The latter is only needed
when a global estimation is required. Hence, errors produced by inaccurate
fusion do not carry over across time steps. This is advantageous in
a number of situations where fusion is not needed or cannot be made
at each time step. If global structural data is exactly known at each
node, the stability of local filters is equivalent to that of the
centralized Kalman filter. Otherwise, we give conditions to guarantee
stability and bound the estimation error induced by inaccurate global
structural data fusion. We also present numerical experiments showing
the advantage of our method over other available alternatives.

\appendix

\section{Proofs}
\begin{pf}
{[}of Proposition~\ref{prop:rim-dist-1}{]} Let $\sigma_{n}(X)$
and $\lambda_{n}(X)$ denote the singular values and eigenvalues of
matrix $X$, respectively. We have
\begin{align*}
\delta\left(P,Q\right) & =\left\Vert \log Q^{-1/2}PQ^{-1/2}\right\Vert _{\mathrm{F}}\\
 & =\sqrt{\sum_{n=1}^{N}\sigma_{n}^{2}\left(\log Q^{-1/2}PQ^{-1/2}\right)}\\
 & =\sqrt{\sum_{n=1}^{N}\lambda_{n}^{2}\left(\log Q^{-1/2}PQ^{-1/2}\right)}\\
 & =\sqrt{\sum_{n=1}^{N}\log^{2}\lambda_{n}\left(Q^{-1/2}PQ^{-1/2}\right)}\\
 & =\sqrt{\sum_{n=1}^{N}\log^{2}\lambda_{n}\left(PQ^{-1}\right)}.
\end{align*}
Then, $\delta\left(P,Q\right)$ equals the distance defined in~\cite[Definition 1.4]{bougerol1993kalman}.
Hence, Properties~1 and~3 follow from~\cite{bougerol1993kalman},
and Property~2 follows from~\cite[Proposition 6]{sui2018accuracy}.

For Property~4 we have
\begin{align*}
\delta\left(P+R,P\right)= & \sqrt{\sum_{n=1}^{N}\log^{2}\lambda_{n}\left(P^{-1/2}(P+R)P^{-1/2}\right)}\\
= & \sqrt{\sum_{n=1}^{N}\log^{2}\lambda_{n}\left(I+P^{-1/2}RP^{-1/2}\right)}\\
= & \sqrt{\sum_{n=1}^{N}\log^{2}\left(1+\lambda_{n}\left(P^{-1/2}RP^{-1/2}\right)\right)}
\end{align*}
Then
\begin{align*}
 & \delta\left(P+R,P\right)\\
\leq & \sqrt{\sum_{n=1}^{N}\log^{2}\left(1-\left|\lambda_{n}\left(P^{-1/2}RP^{-1/2}\right)\right|\right)}\\
\leq & \sqrt{N}\max_{n}\left|\log\left(1-\left|\lambda_{n}\left(P^{-1/2}RP^{-1/2}\right)\right|\right)\right|\\
= & \sqrt{N}\left|\log\left(1-\max_{n}\left|\lambda_{n}\left(P^{-1/2}RP^{-1/2}\right)\right|\right)\right|\\
\leq & \sqrt{N}\left|\log\left(1-\left\Vert P^{-1/2}RP^{-1/2}\right\Vert \right)\right|\\
\leq & \sqrt{N}\left|\log\left(1-\left\Vert P^{-1}\right\Vert \left\Vert R\right\Vert \right)\right|.
\end{align*}
$\hfill\qed$
\end{pf}
\begin{pf}
{[}of Lemma~\ref{lem:autocorr}{]} From~\cite[Proposition 1.1]{Lance1995},
\begin{align*}
\mathcal{E}\left\{ xy^{\top}\right\} \mathcal{E}\left\{ yx^{\top}\right\}  & \leq\left\Vert \mathcal{E}\left\{ yy^{\top}\right\} \right\Vert \mathcal{E}\left\{ xx^{\top}\right\} .
\end{align*}
Then
\begin{align*}
\left\Vert \mathcal{E}\left\{ xy^{\top}\right\} \right\Vert ^{2} & =\left\Vert \mathcal{E}\left\{ xy^{\top}\right\} \mathcal{E}\left\{ yx^{\top}\right\} \right\Vert \\
 & \leq\left\Vert \mathcal{E}\left\{ xx^{\top}\right\} \right\Vert \left\Vert \mathcal{E}\left\{ yy^{\top}\right\} \right\Vert ,
\end{align*}
and the result follows.$\hfill\qed$
\end{pf}
\begin{pf}
{[}of Lemma~\ref{lem:LTV-bound}{]} We have
\begin{align*}
\hat{x}_{t} & =\hat{A}_{t-1}\hat{x}_{t-1}+u_{t}\\
 & =A_{t-1}\hat{x}_{t-1}+\left(\hat{A}_{t-1}-A_{t-1}\right)\hat{x}_{t-1}+u_{t}\\
 & =\sum_{s=1}^{t}\Pi_{t,s}\left[\tilde{A}_{s-1}\hat{x}_{s-1}+u_{s}\right],
\end{align*}
where $\Pi_{t,s}=A_{t-1}\times\cdots\times A_{s}$. Then
\[
\mathcal{E}\left\{ x_{t}x_{t}^{\top}\right\} =R_{t}^{(1)}+R_{t}^{(2)}+R_{t}^{(3)}+R_{t}^{(4)},
\]
with
\begin{align*}
R_{t}^{(1)} & =\sum_{s,r=1}^{t}\Pi_{t,s}\tilde{A}_{s-1}\mathcal{E}\left\{ \hat{x}_{s-1}\hat{x}_{r-1}^{\top}\right\} \tilde{A}_{r-1}^{\top}\Pi_{t,r}^{\top},\\
R_{t}^{(2)} & =\sum_{s,r=1}^{t}\Pi_{t,s}\tilde{A}_{s-1}\mathcal{E}\left\{ \hat{x}_{s-1}u_{r}^{\top}\right\} \Pi_{t,r}^{\top},\\
R_{t}^{(3)} & =\left(R_{t}^{(2)}\right)^{\top},\\
R_{t}^{(4)} & =\sum_{s,r=1}^{t}\Pi_{t,s}\mathcal{E}\left\{ u_{s}u_{r}^{\top}\right\} \Pi_{t,r}^{\top}.
\end{align*}

Let $\nu_{t-1}=\max_{1\leq s\leq t-1}\left\Vert \hat{x}_{t}\right\Vert _{\star}$.
Then
\begin{align*}
 & \left\Vert R_{t}^{(1)}\right\Vert \\
\leq & \sum_{s,r=1}^{t}\left\Vert \Pi_{t,s}\right\Vert \left\Vert \tilde{A}_{s-1}\right\Vert \left\Vert \left\langle x_{r-1}^{\top},x_{s-1}\right\rangle _{\star}\right\Vert \left\Vert \tilde{A}_{r-1}^{\top}\right\Vert \left\Vert \Pi_{t,r}^{\top}\right\Vert \\
\leq & \mu^{2}\sum_{s,r=1}^{t}\left\Vert \Pi_{t,s}\right\Vert \left\Vert \Pi_{t,r}^{\top}\right\Vert \left\Vert x_{s-1}\right\Vert _{\star}\left\Vert x_{r-1}\right\Vert _{\star}\\
\leq & \mu^{2}\left(\sum_{s=1}^{t}\left\Vert \Pi_{t,s}\right\Vert \right)^{2}\nu_{t-1}^{2}\leq\mu^{2}\gamma^{2}\nu_{t-1}^{2}.
\end{align*}
Also
\begin{align*}
\left\Vert R_{t}^{(2)}\right\Vert  & =\left\Vert R_{t}^{(3)}\right\Vert \\
 & \leq\sum_{s,r=1}^{t}\left\Vert \Pi_{t,s}\right\Vert \left\Vert \tilde{A}_{s-1}\right\Vert \left\Vert \left\langle u_{r}^{\top},x_{s-1}\right\rangle _{\star}\right\Vert \left\Vert \Pi_{t,r}^{\top}\right\Vert \\
 & \leq\mu\sum_{s,r=1}^{t}\left\Vert \Pi_{B}\left(t,s\right)\right\Vert \left\Vert x_{s-1}\right\Vert _{\star}\left\Vert u_{s}\right\Vert _{\star}\left\Vert \Pi_{B}^{\top}\left(t,r\right)\right\Vert \\
 & \leq\mu\gamma^{2}\bar{u}\nu_{t-1},
\end{align*}
and
\[
\left\Vert R_{t}^{(4)}\right\Vert \leq\sum_{s,r=1}^{t}\left\Vert \Pi_{t,s}\right\Vert \left\Vert u_{s}\right\Vert _{\star}^{2}\left\Vert \Pi_{t,r}^{\top}\right\Vert \leq\gamma^{2}\bar{u}^{2}.
\]
We then obtain
\begin{multline*}
\left\Vert x_{t}\right\Vert _{\star}=\left\Vert \mathcal{E}\left\{ x_{t}x_{t}^{\top}\right\} \right\Vert ^{1/2}\\
\leq\sqrt{\left\Vert R_{t}^{(1)}\right\Vert +2\left\Vert R_{t}^{(2)}\right\Vert +\left\Vert R_{t}^{(4)}\right\Vert }=\gamma\mu\nu_{t-1}+\gamma\bar{u}.
\end{multline*}
Since $x_{0}=0$, it follows that $\nu_{t}\leq\gamma\mu\nu_{t-1}+\gamma\bar{u}$.
Hence, 
\[
\nu_{t}\leq\frac{\gamma\bar{u}}{1-\gamma\mu},
\]
and the result follows.$\hfill\qed$
\end{pf}
\begin{pf}
{[}of Lemma~\ref{lem:riemmanian}{]} We have
\begin{align}
\delta\left(\Sigma_{t|t}^{i},\Sigma_{t|t}\right) & =\delta\left(\Sigma_{t|t-1}^{-i}+\Psi_{t}^{i},\Sigma_{t|t-1}^{-1}+\Psi_{t}\right)\nonumber \\
 & =\delta\left(\Sigma_{t|t-1}^{-i}+\Psi_{t}^{i},\Sigma_{t|t-1}^{-1}+\Psi_{t}^{i}\right)\nonumber \\
 & +\delta\left(\Sigma_{t|t-1}^{-1}+\Psi_{t}^{i},\Sigma_{t|t-1}^{-1}+\Psi_{t}\right).\label{eq:aux1}
\end{align}

Since $\left\Vert \tilde{\Psi}_{t}^{i}\right\Vert \leq\left\Vert \Sigma_{t|t}\right\Vert ^{-1}$,
we have from Proposition~\ref{prop:rim-dist-1}~4 that
\begin{align}
 & \delta\left(\Sigma_{t|t-1}^{-1}+\Psi_{t}^{i},\Sigma_{t|t-1}^{-1}+\Psi_{t}\right)\nonumber \\
= & \delta\left(\Sigma_{t|t-1}^{-1}+\Psi_{t}+\tilde{\Psi}_{t}^{i},\Sigma_{t|t-1}^{-1}+\Psi_{t}\right)\nonumber \\
= & \delta\left(\Sigma_{t|t}^{-1}+\tilde{\Psi}_{t}^{i},\Sigma_{t|t}^{-1}\right)\nonumber \\
\leq & \sqrt{N}\left|\log\left(1-\left\Vert \Sigma_{t|t}\right\Vert \left\Vert \tilde{\Psi}_{t}^{i}\right\Vert \right)\right|.\label{eq:aux2}
\end{align}
Also
\begin{align}
 & \delta\left(\Sigma_{t|t-1}^{-i}+\Psi_{t}^{i},\Sigma_{t|t-1}^{-1}+\Psi_{t}^{i}\right)\nonumber \\
\leq & \delta\left(\Sigma_{t|t-1}^{-i},\Sigma_{t|t-1}^{-1}\right)\nonumber \\
= & \delta\left(\Sigma_{t|t-1}^{i},\Sigma_{t|t-1}\right)\nonumber \\
= & \delta\left(A\Sigma_{t-1|t-1}^{i}A^{\top}+Q,A\Sigma_{t-1|t-1}A^{\top}+Q\right)\nonumber \\
\leq & \lambda_{t}\delta\left(\Sigma_{t-1|t-1}^{i},\Sigma_{t-1|t-1}\right),\label{eq:aux3}
\end{align}
with
\begin{align*}
\lambda_{t} & =\frac{\alpha_{t}}{\alpha_{t}+\beta_{t}},\\
\alpha_{t} & =\max\left\{ \left\Vert A\Sigma_{t-1|t-1}^{i}A^{\top}\right\Vert ,\left\Vert A\Sigma_{t-1|t-1}A^{\top}\right\Vert \right\} ,\\
\beta_{t} & =\left\Vert Q^{-1}\right\Vert ^{-1}.
\end{align*}

Now
\begin{align*}
\alpha_{t} & \leq\left\Vert A\right\Vert ^{2}\max\left\{ \left\Vert \Sigma_{t-1|t-1}^{i}\right\Vert ,\left\Vert \Sigma_{t-1|t-1}\right\Vert \right\} \\
 & \leq\left\Vert A\right\Vert ^{2}\left(\left\Vert \Sigma_{t-1|t-1}\right\Vert +\left\Vert \Sigma_{t-1|t-1}-\Sigma_{t-1|t-1}^{i}\right\Vert \right)\\
 & \leq\left\Vert A\right\Vert ^{2}\left\Vert \Sigma_{t-1|t-1}\right\Vert e^{\delta\left(\Sigma_{t-1|t-1},\Sigma_{t-1|t-1}^{i}\right)}
\end{align*}
We then have
\begin{equation}
\lambda_{t}\leq\frac{\left\Vert A\right\Vert ^{2}\left\Vert \Sigma_{t-1|t-1}\right\Vert }{\left\Vert A\right\Vert ^{2}\left\Vert \Sigma_{t-1|t-1}\right\Vert +\left\Vert Q^{-1}\right\Vert ^{-1}e^{-\delta\left(\Sigma_{t-1|t-1},\Sigma_{t-1|t-1}^{i}\right)}}.\label{eq:aux4}
\end{equation}
The result then follows by putting~(\ref{eq:aux4}) into~(\ref{eq:aux3})
and the resulting equation, together with~(\ref{eq:aux2}) into~(\ref{eq:aux1}).$\hfill\qed$
\end{pf}
\begin{pf}
{[}of Lemma~\ref{lem:xi}{]} From~(\ref{eq:est-update})-(\ref{eq:est-prediction}),
we have
\begin{align*}
\breve{\xi}_{t|t}^{i} & =\left(I-\Phi_{t}\right)A\breve{\xi}_{t-1|t-1}^{i}+\Sigma_{t|t}C_{t}^{i\top}R_{t}^{-i}y_{t}^{i},\\
\xi_{t|t}^{i} & =\left(I-\Phi_{t}^{i}\right)A\xi_{t-1|t-1}^{i}+\Sigma_{t|t}^{i}C_{t}^{i\top}R_{t}^{-i}y_{t}^{i}.
\end{align*}
Then
\begin{align*}
\tilde{\xi}_{t|t}^{i} & =\left(I-\Phi_{t}^{i}\right)A\xi_{t-1|t-1}^{i}-\left(I-\Phi_{t}\right)A\breve{\xi}_{t-1|t-1}^{i}\\
 & +\left[\Sigma_{t|t}^{i}-\Sigma_{t|t}\right]C_{t}^{i\top}R_{t}^{-i}y_{t}^{i}\\
 & =\left(I-\Phi_{t}-\tilde{\Phi}_{t}^{i}\right)A\tilde{\xi}_{t-1|t-1}^{i}-\tilde{\Phi}_{t}\breve{\xi}_{t|t-1}^{i}+\tilde{\Sigma}_{t|t}^{i}\mathring{\psi}_{t}^{i}\\
 & =\left(I-\Phi_{t}-\tilde{\Phi}_{t}^{i}\right)A\tilde{\xi}_{t-1|t-1}^{i}+\left[\begin{array}{cc}
\tilde{\Phi}_{t}^{i}, & \tilde{\Sigma}_{t|t}^{i}\end{array}\right]\mathfrak{y}_{t}^{i},
\end{align*}
where
\begin{align*}
\tilde{\Phi}_{t}^{i} & =\Sigma_{t|t}^{i}\Psi_{t}^{i}-\Sigma_{t|t}\Psi_{t}=\tilde{\Sigma}_{t|t}^{i}\Psi_{t}+\Sigma_{t|t}\tilde{\Psi}_{t}^{i}+\tilde{\Sigma}_{t|t}^{i}\tilde{\Psi}_{t}^{i}.
\end{align*}

Now
\[
\left\Vert \tilde{\Sigma}_{t|t}^{i}\right\Vert \leq\left(e^{\delta\left(\Sigma_{t|t}^{i},\Sigma_{t|t}\right)}-1\right)\left\Vert \Sigma_{t|t}\right\Vert .
\]
Hence
\begin{align*}
\left\Vert \tilde{\Phi}_{t}^{i}\right\Vert  & \leq\left\Vert \tilde{\Sigma}_{t|t}^{i}\right\Vert \left\Vert \Psi_{t}\right\Vert +\left\Vert \Sigma_{t|t}\right\Vert \left\Vert \tilde{\Psi}_{t}^{i}\right\Vert +\left\Vert \tilde{\Sigma}_{t|t}^{i}\right\Vert \left\Vert \tilde{\Psi}_{t}^{i}\right\Vert \\
 & \leq\left[\left(e^{\delta\left(\Sigma_{t|t}^{i},\Sigma_{t|t}\right)}-1\right)\left\Vert \Psi_{t}\right\Vert +\right.\\
 & +\left.e^{\delta\left(\Sigma_{t|t}^{i},\Sigma_{t|t}\right)}\left\Vert \tilde{\Psi}_{t}^{i}\right\Vert \right]\left\Vert \Sigma_{t|t}\right\Vert .
\end{align*}
$\hfill\qed$
\end{pf}

\begin{pf}
{[}of Lemma~\ref{lem:xi-bound}{]} It follows from~(\ref{eq:stab-cond})
that $\bar{\tilde{\psi}}<\bar{\sigma}^{-1}$, which in turn implies
the condition of Lemma~\ref{lem:riemmanian}. From the latter we
then obtain
\begin{multline}
\delta\left(\Sigma_{t|t}^{i},\Sigma_{t|t}\right)\leq\\
\frac{\delta\left(\Sigma_{t-1|t-1}^{i},\Sigma_{t-1|t-1}\right)}{1+\frac{\left\Vert Q^{-1}\right\Vert ^{-1}}{\bar{\sigma}\left\Vert A\right\Vert ^{2}}e^{-\delta\left(\Sigma_{t-1|t-1}^{i},\Sigma_{t-1|t-1}\right)}}+\bar{\upsilon}\left(\bar{\tilde{\psi}}\right).\label{eq:delta-it}
\end{multline}
Since $\delta\left(\Sigma_{1|1}^{i},\Sigma_{1|1}\right)=0$, we have
from Lemma~\ref{lem:implicit} that the iterations~(\ref{eq:delta-it})
converge to $\bar{\delta}\left(\bar{\tilde{\psi}}\right)$, and 
\begin{equation}
\delta\left(\Sigma_{t|t}^{i},\Sigma_{t|t}\right)\leq\bar{\delta}\left(\bar{\tilde{\psi}}\right).\label{eq:aux-2}
\end{equation}
Using~(\ref{eq:aux-2}) in Lemma~\ref{lem:xi} we obtain
\begin{align*}
\left\Vert \tilde{\Phi}_{t}^{i}\right\Vert  & \leq\left[\left(e^{\bar{\delta}\left(\bar{\tilde{\psi}}\right)}-1\right)\bar{\psi}+e^{\bar{\delta}\left(\bar{\tilde{\psi}}\right)}\bar{\tilde{\psi}}\right]\bar{\sigma}=\bar{\phi}\left(\bar{\tilde{\psi}}\right),\\
\left\Vert \tilde{\Sigma}_{t|t}^{i}\right\Vert  & \leq\left(e^{\bar{\delta}\left(\bar{\tilde{\psi}}\right)}-1\right)\bar{\sigma}=\bar{\tilde{\sigma}}\left(\bar{\tilde{\psi}}\right).
\end{align*}

Let $\mathfrak{u}_{t}^{i}=\left[\begin{array}{cc}
\tilde{\Phi}_{t}^{i}, & \tilde{\Sigma}_{t|t}^{i}\end{array}\right]\mathfrak{y}_{t}^{i}$. We have
\[
\mathcal{E}\left\{ \mathfrak{u}_{t}^{i}\mathfrak{u}_{t}^{i\top}\right\} =\left[\begin{array}{cc}
\tilde{\Phi}_{t}^{i}, & \tilde{\Sigma}_{t|t}^{i}\end{array}\right]\mathcal{E}\left\{ \mathfrak{y}_{t}^{i}\mathfrak{y}_{t}^{i\top}\right\} \left[\begin{array}{c}
\tilde{\Phi}_{t}^{i}\\
\tilde{\Sigma}_{t|t}^{i}
\end{array}\right].
\]
It then follows that
\begin{align*}
\left\Vert \mathcal{E}\left\{ \mathfrak{u}_{t}^{i}\mathfrak{u}_{t}^{i\top}\right\} \right\Vert  & \leq\left\Vert \mathcal{E}\left\{ \mathfrak{y}_{t}^{i}\mathfrak{y}_{t}^{i\top}\right\} \right\Vert \left(\left\Vert \tilde{\Phi}_{t}^{i}\right\Vert ^{2}+\left\Vert \tilde{\Sigma}_{t|t}^{i}\right\Vert ^{2}\right)\\
 & \leq\bar{\mathfrak{y}}^{2}\left(\bar{\phi}^{2}\left(\bar{\tilde{\psi}}\right)+\bar{\tilde{\sigma}}^{2}\left(\bar{\tilde{\psi}}\right)\right),
\end{align*}
or $\left\Vert \mathfrak{u}_{t}^{i}\right\Vert _{\star}\leq\bar{\mathfrak{y}}\sqrt{\bar{\phi}^{2}\left(\bar{\tilde{\psi}}\right)+\bar{\tilde{\sigma}}^{2}\left(\bar{\tilde{\psi}}\right)}$.

Equation~(\ref{eq:perturbedLTV}) defines a perturbed linear system
with $-\tilde{\Phi}_{t}^{i}A$ being the perturbation of the nominal
state-transition matrix $\left(I-\Phi_{t}\right)A$ and $\mathfrak{u}_{t}^{i}$
being the input. Since $\bar{\tilde{\psi}}\bar{\gamma}\left\Vert A\right\Vert <1,$
we can apply Lemma~\ref{lem:LTV-bound} to this perturbed system
to obtain~(\ref{eq:xi-bound}).$\hfill\qed$
\end{pf}
\bibliographystyle{unsrt}
\bibliography{references}

\end{document}